\documentstyle[aps,psfig]{revtex}
\begin{document}
\title{Generation of phase-coherent states}
\author{Giacomo M. D'Ariano, Matteo G. A. Paris and Massimiliano F. Sacchi}
\address{{\sc Theoretical Quantum Optics Group}\thanks{\sc
http://enterprise.pv.infn.it}\\
Dipartimento di Fisica 'Alessandro Volta' dell'Universit\`a di Pavia \\
Istituto Nazionale di Fisica della Materia -- Unit\`a di Pavia \\
{via Bassi 6 -- I-27100 Pavia, ITALY}}
\date{\today}\maketitle
\begin{abstract}
An interaction scheme involving nonlinear $\chi^{(2)}$ media is suggested
for the generation of phase-coherent states (PCS). The setup is based on
parametric amplification of vacuum followed by up-conversion of the resulting
twin-beam. The involved nonlinear interactions are studied by the exact
numerical diagonalization. An experimentally achievable working regime to
approximate PCS with high conversion rate is given, and the validity of
parametric approximation is discussed.
\end{abstract}
\section{Introduction}\label{s:intro}
Optical processes taking place in $\chi^{(2)}$ media involve three light
waves, yielding to a considerably rich variety of nonlinear phenomena,
both in the semi-classical \cite{blo} and in the quantum domain \cite{man}.
The quantum statistical properties of the radiation coming out from such
interactions have attracted much attention (see for example Refs.
\cite{yue,yam}).
Squeezing, anti-bunching and entanglement have been predicted and subsequently
observed in a series of fascinating experiments \cite{spe,qts}. Most of the
theoretical approaches to a quantum theory of three-waves devices have
been carried out using the so-called parametric approximation \cite{mol}.
In this framework one of the field modes is in a strong semi-classical
coherent state, so that its depletion as well as its quantum fluctuations
can be neglected.
A fully analytical treatment of the quantum dynamics is not
available, whereas numerical methods have been developed in the cases of
photon number \cite{old} or coherent input states \cite{igo}.
In this paper the three-wave dynamics is evaluated without approximation
for arbitrary input states, resorting to the numerical block-diagonalization
of the Hamiltonian in invariant subspaces of the constants of motion.
\par In the rotating wave approximation the non-degenerate three-wave
interactions are described by the Hamiltonian
\begin{eqnarray}
\hat H \propto \chi^{(2)} \left[ a b c^{\dag} + a^{\dag} b^{\dag} c \right]
\label{ham}\:,
\end{eqnarray}
where $a$, $b$ and $c$ are the annihilation operators of the three relevant
modes, whose frequencies satisfy the relation $\omega_c =\omega_a +
\omega_b$. Depending on the input state of the field, the
Hamiltonian (\ref{ham}) describes phase-insensitive amplification or
frequency up- or down-conversion.
The first kind of process occurs in situations with small $a$ and $b$
and large coherent $c$, so that the pumping mode can be considered as
undepleted and treated as a c-number, in the so-called parametric
approximation. On the other hand, when all the three modes participates
to the quantum dynamics we are in the presence of frequency up- and
down-conversion processes. \par
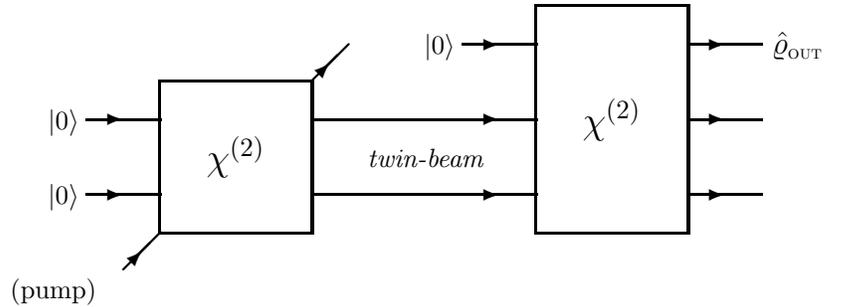
\begin{figure}[h]
\hspace{-30pt}
\begin{center}
\begin{picture}(12,6)(-2,-2)
  \setlength{\unitlength}{1cm}
\thicklines
\put(0,0){\framebox(2,2){\Large $\chi^{(2)}$}}
 \put(5,0){\framebox(2,3){\Large $\chi^{(2)}$}}
\put(2.75,.86){{\em twin-beam}} \put(3.5,2.375){$\vert 0\rangle$}
\put(8.125,2.375){\large $\hat\varrho_{\hbox{\scriptsize\sc out}}$}
\put(-2,-.875){(pump)} \put(-1.5,.375){$\vert 0\rangle$}
\put(-1.5,1.375){$\vert 0\rangle$} \multiput(-.5,.5)(0,1){2}{\vector(1,0){0}}
\multiput(4.5,.5)(0,1){3}{\vector(1,0){0}} \multiput(7.5,.5)(0,1){3}{\vector(1,0){0}}
\multiput(-.25,-.25)(2.5,2.5){2}{\vector(1,1){0}}
\multiput(-1,.5)(0,1){2}{\line(1,0){1}} \multiput(2,.5)(0,1){2}{\line(1,0){2}}
\multiput(4,.5)(0,1){3}{\line(1,0){1}} \multiput(7,.5)(0,1){3}{\line(1,0){1}}
\multiput(-.5,-.5)(2.5,2.5){2}{\line(1,1){.5}}
\end{picture}
\end{center}
\vspace{20pt}
\caption{Scheme of generation of phase-coherent states.}
\label{f:fig1}
\end{figure}

In the present paper we are interested in the situation depicted in
Fig. \ref{f:fig1}, where the interaction Hamiltonian (\ref{ham}) is
applied twice: in the first step as a parametric (spontaneous)
down-conversion of the vacuum state, generating a twin-beam on modes $a$
and $b$, and in the second step as the up-conversion of twin-beam into
mode $c$.
\par\noindent
This scheme is of interest because, as shown in the following, the outgoing
quantum state of radiation closely resembles the phase-coherent state (PCS)
\begin{eqnarray}
\vert\lambda\rangle = \sqrt{1-\vert\lambda\vert^2}\: \sum_{n=0}^{\infty }\:
\lambda ^n\:\vert
n\rangle\label{pcsdef}\:,
\end{eqnarray}
which has been introduced by Shapiro {\em et al} in Ref. \cite{sha}.
The phase-coherent states are interesting because they are optimum phase
states for both the S\"ussmann and the reciprocal peak likelihood
\cite{sha} measure of phase uncertainty \cite{bia,dae}. On the other
hand, they also could serve as {\em seed }
state \cite{buz,wod,ole} in sampling canonical phase distribution by
unconventional heterodyne detection \cite{sac}. Moreover, one should mention
that the PCS maintain phase coherence under phase amplification \cite{pha},
such that they are privileged states for phase-based communication channels.
\par
In suggesting the present scheme we have been inspired by
Ref. \cite{rev}, where
an ideal scheme using a photon number duplicator (PND) was suggested for
PCS synthesis from twin-beam. As a matter of fact, in such photon recombination
process the PND is well approximated by the up-conversion from
Hamiltonian (\ref{ham})
with mode $c$ initially in the vacuum \cite{mac,mos}. In this paper we
show that the
interaction scheme sketched in Fig. \ref{f:fig1} is indeed effective for
the generation of PCS. \par
The paper is structured as follows. In Section \ref{s:thr} we briefly
describe our approach to the evaluation of the dynamics of the three-wave
interactions,
and discuss the validity of the parametric approximation in the generation of
twin-beam state. In Section \ref{s:upc} we analyze the performances of
the twin-beam up-conversion in producing phase-coherent states in the second
stage
of the scheme of Fig. \ref{f:fig1}.
Finally, Section \ref{s:outro} closes the paper with some concluding remarks.
\section{Dynamics of the three-wave interactions}\label{s:thr}
The overall input state describing the three involved modes can be written
in the Fock basis as
\begin{eqnarray}
\vert\psi_0\rangle=\sum_{n_1,n_2,n_3}^{N}c_{n_1,n_2,n_3}\:\vert
n_1,n_2,n_3\rangle
\label{gen}\:,
\end{eqnarray}
where $N$ is an arbitrarily large integer, which denotes the largest
not-negligible Fock component.
In order to compute the dynamical evolution of $\vert\psi\rangle$
\begin{eqnarray}
\vert\psi_t\rangle=\exp\left(-it \hat H\right)\:\vert\psi_0\rangle
\label{evol}\:,
\end{eqnarray}
one should, in principle, diagonalize the full Hamiltonian matrix in the Fock
basis. This becomes a very difficult task when the truncation $N$ of the
Fock space increases, becoming unrealistic for $N$ exceeding few teens.
However, one can notice that the Hamiltonian (\ref{ham}) admits two
independent constants of motion. For the sake of convenience we choose them
as the following ones
\begin{eqnarray}
\hat S = \frac{1}{2}\left[ a^{\dag} a+ b^{\dag} b +
2 c^{\dag} c \right]\:,  \quad\quad
\hat K =a^{\dag} a + c^{\dag} c
\label{kost}\:.
\end{eqnarray}
Conservation of $\hat S$ and $\hat K$ means that subspaces
corresponding to given eigenvalues
of these quantities are invariant under the action of the Hamiltonian
(\ref{ham}), as $[\hat H, \hat S]=0$ and $[\hat H, \hat K]=0$. In other words,
the Hilbert space ${\cal H}_a \otimes {\cal H}_b \otimes {\cal H}_c$ can
be decomposed into the direct sum of subspaces that are invariant under the
action
of the unitary evolution operator in Eq. (\ref{evol}). Such a decomposition can
be written as follows
\begin{eqnarray}
{\cal H}_a \otimes {\cal H}_b \otimes {\cal H}_c =\oplus_{s=0}^{N}
\oplus_{k=0}^{s} {\cal H}_{sk}
\label{decbs1}\;,
\end{eqnarray}
where
\begin{eqnarray}
{\cal H}_{sk}=\hbox{Span}\left\{\:\vert k-n\rangle \otimes \vert s-k-n\rangle
\otimes \vert n\rangle \right\}\nonumber \\ n\in [0,min(k,s-k)]
\label{decbs2}\;,
\end{eqnarray}
$\hbox{Span}\{\cdot\}$ denoting the Hilbert space linearly spanned by the
orthogonal
vectors within the brackets, and $\vert n_1\rangle \otimes \vert n_2\rangle
\otimes
\vert n_3\rangle\equiv \vert n_1,n_2,n_3\rangle$ representing the state which is
simultaneously eigenvector of the number operator of the three modes.
The Hamiltonian (\ref{ham}) can be consistently rewritten as a follows
\begin{eqnarray}
\hat H= \sum_{sk}  \hat h_{sk}
\label{decham}\;,
\end{eqnarray}
with $\hat h_{sk}$ acting on ${\cal H}_{sk}$ only. Correspondingly, the
representation for $\vert\psi_0\rangle$ can be written in the following fashion
\begin{eqnarray}
\vert\psi_0\rangle= \sum_{s=0}^{N}\sum_{k=0}^{s}\!\sum_{n=0}^{min(k,s-k)}
\!\!\!\!\!c_{k-n,s-k-n,n}\:\vert k-n,s-k-n,n\rangle
\label{spe}\:,
\end{eqnarray}
which enlightens the invariant subspaces structure. In this way
one needs to diagonalize the Hamiltonian only {\em inside}
each invariant subspace, thus leading to a considerable saving of resources.
\par
As a first application of the above method we look for the conditions under
which the
parametric approximation is justified in describing the process of
frequency down-conversion between a strong semi-classical pump and the
vacuum. The input state is given by $\vert\psi_0\rangle=\vert 0,0,\alpha
\rangle$, $\alpha >> 1$ being the
amplitude of a considerably excited coherent state.
In the parametric approximation the pump mode $c$ in the
Hamiltonian (\ref{ham}) is replaced by the c-number $\alpha$, thus
neglecting its quantum fluctuations as well its depletion.
Within such approximation the dynamics of the input state
$\vert 0,0\rangle$ is governed by the evolution operator
\begin{eqnarray}
\hat U = \exp\left\{ \zeta a^{\dag} b^{\dag} - \bar\zeta a b \right\}
\label{pia}\:,
\end{eqnarray}
where $\zeta=-i\kappa t \alpha$, $t$ being the interaction time and $\kappa$
the coupling constant containing the nonlinear susceptibility.
The evolution governed by Eq. (\ref{pia}) can be easily computed by
means of the Baker-Haussdorff-Campbell formula for the $SU(1,1)$ Lie
algebra \cite{rev,sch,tru}, and
the output is represented by the twin-beam state
\begin{eqnarray}
\vert\chi \rangle = \sqrt{1-\vert\chi\vert^2}\:\sum_{n=0}^{\infty}\:\chi^n\:
\vert n,n \rangle
\label{twb1}\:,
\end{eqnarray}
where
\begin{eqnarray}
\chi = - i \tanh (\kappa t\vert\alpha\vert ) e^{i\phi_{\alpha}}
\label{twb2}\:.
\end{eqnarray}
\par
\begin{figure}[h]
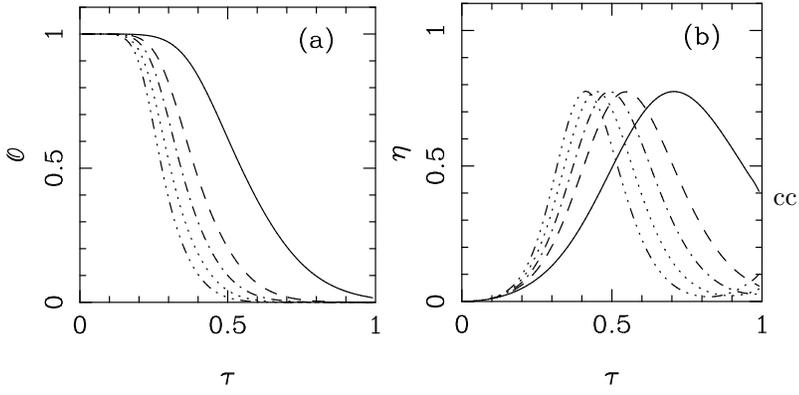

\begin{tabular}{cc}
\psfig{file=fig2_a.ps,width=5cm} &
\psfig{file=fig2_b.ps,width=5cm}
\end{tabular}{cc}
\caption{In (a): overlap ${\cal O}=\sqrt{\langle\chi\vert
\hat\varrho '\vert\chi\rangle}$ between the state $\hat\varrho '$
coming from the exact evolution and the twin-beam $\vert\chi\rangle$
expected within the parametric approximation, as a function of the
scaled time $\tau=\kappa t$ for different values of the pump input
photon number. In (b): energy
conversion rate $\eta$ as a function of the scaled time for different
values of the pump input photon number.
In both plots different line styles denotes different values of pump
intensity: $\langle \hat n_c\rangle=81$ (dot-dot-dashed), $\langle \hat n_c
\rangle=64$ (dotted), $\langle \hat n_c\rangle=49$ (dot-dashed),
$\langle \hat n_c\rangle=36$ (dashed), $\langle \hat n_c\rangle=16$ (solid).
The interaction time leading to maximum conversion rate follows the
relation $\tau_{opt} \propto \langle\hat n_c\rangle^{-1/3} $.}
\label{f:fig2}
\end{figure}

\noindent
In order to check the theoretical results predicted by the parametric
approximation we consider the overlap
$${\cal O}=\sqrt{\langle\chi\vert \hat\varrho ' \vert\chi\rangle}$$
between the state $\hat\varrho '$ coming from the exact evolution
and the expected twin-beam $\vert\chi\rangle$. In Fig. \ref{f:fig2}(a) we show
the behavior of the overlap as a function of the scaled interaction time
$\tau =\kappa t$ for different values of the pump input power.
In order to evaluate the efficiency of the process we
also consider the energy conversion rate $\eta$, which is defined as
\begin{eqnarray}
\eta= \frac{1}{2}\frac{\hbox{Tr}\left[\hat\varrho '\:
(\hat n_a + \hat n_b)\right]}
{\hbox{Tr}\left[\hat\varrho^{in}\hat n_c\right]}
\label{conv}\:,
\end{eqnarray}
where $\hat\varrho^{in}=\vert\psi_0\rangle\langle\psi_0\vert$.
In Eq. (\ref{conv}) $\eta$ runs between zero and one, the factor $1/2$
coming from frequency conversion. In Fig. \ref{f:fig2}(b) we show the
behavior of $\eta$ as a function of the scaled interaction time $\tau$
for different values of the input power, as in Fig. \ref{f:fig2}(a).
It is apparent that parametric approximation is valid
also for moderate input power, and that one has a considerably wide range of
values of the interaction time leading to an overlap very close to unit,
the weaker is the pump, the larger is this range. On the other hand, these
values of the interaction time correspond to a low conversion rate.
By the way, we note that the interaction time leading to
maximum conversion rate follows the relation $\tau_{opt} \propto \langle\hat
n_c\rangle^{-1/3} $.
\section{Twin-beam up conversion}\label{s:upc}
\begin{figure}[h]
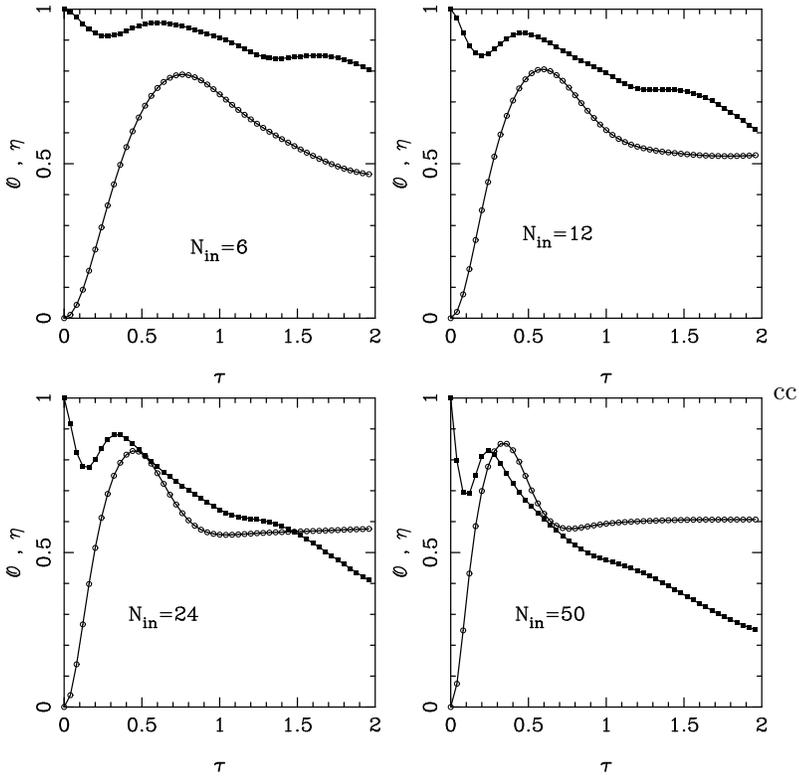

\begin{tabular}{cc}
\psfig{file=fig3_a.ps,width=5cm} &
\psfig{file=fig3_b.ps,width=5cm} \\
\psfig{file=fig3_c.ps,width=5cm} &
\psfig{file=fig3_d.ps,width=5cm}
\end{tabular}{cc}
\caption{Behavior of ${\cal O}$ and $\eta$ as a
function of the scaled interaction time for different intensity of the
incoming twin-beam.}
\label{f:fig3}
\end{figure}

In this section we analyze the second step of the PCS generation setup
reported in Fig.\ref{f:fig1}, namely the three-wave interaction
starting from the twin-beam input state
\begin{eqnarray}
\vert\chi\rangle = \sqrt{1-\vert\chi\vert^2}\:\sum_{n=0}^{\infty }
\:\chi^n\:|n,n,0\rangle
\label{twbin}\:.
\end{eqnarray}
The complex amplitude $\chi$ is confined in the unit circle, and the mean
photon number pertaining the state (\ref{twbin}) is given by
$$\langle\chi\vert\hat n_a+\hat n_b+\hat n_c\vert \chi\rangle =
2\vert\chi\vert^2/(1-\vert\chi\vert^2).$$
The synthesis of the PCS (\ref{pcsdef}) starting from $\vert\chi\rangle$ would
be easily
achieved  by having at disposal a device that performs the photon number
recombination $$ |n,n,0\rangle \longrightarrow |0,0,n\rangle\:.$$
Such kind of transformation has been analyzed in Ref. \cite{rev},
and has been shown to correspond to the interaction Hamiltonian
\begin{eqnarray}
\hat H_r=a^{\dag} b^{\dag} (b^{\dag} b +1)^{-{1\over 2}} c +
c^{\dag}(b^{\dag} b +1)^{-{1\over 2}}ab
\label{pnr}\:.
\end{eqnarray}
Unfortunately, the Hamiltonian (\ref{pnr}) cannot be realized by known optical
devices. However, one may notice that the perfect number recombination
$\vert 1,1,0\rangle\rightarrow\vert 0,0,1\rangle$ is performed by the
Hamiltonian (\ref{ham}), and this suggests to substitute the intensity
dependent factor in Eq. (\ref{pnr}) by its expectation value. In spite of
this rather crude approximation, the trilinear interaction (\ref{ham}) has
been shown \cite{mac,mos} to provide a good approximation of the photon
recombination in the case of a single photon number state at the input.
Here, we analyze the case of the input twin-beam state
(\ref{twbin}). Our aim is to demonstrate that the scheme
of Fig. \ref{f:fig1} is indeed effective in synthesizing a PCS.
As a parameter to evaluate the effectiveness of PCS synthesis we use the
overlap ${\cal O}= \sqrt{\langle\lambda \vert\hat\varrho^{out}
\vert\lambda \rangle}$
between the state
\begin{eqnarray}
\hat\varrho^{out} = \hbox{Tr}_{ab} \left[\exp\left(-it\hat H\right)\:
\vert\chi \rangle\langle
\chi \vert\:\exp\left(it\hat H\right)\right]
\label{out}\:,
\end{eqnarray}
exiting the $\chi^{(2)}$ crystal in the mode $c$
and a theoretical PCS $\vert\lambda \rangle$
corresponding to the same mean photon number. In order to evaluate the
efficiency of the process we also consider the conversion rate $\eta$,
defined as follows
\begin{eqnarray}
\eta= 2 \frac{\hbox{Tr}(\hat\varrho^{out}\:\hat n_c)}{\langle\chi\vert
(\hat n_a + \hat n_b)\vert\chi\rangle}
\label{eta2}\:.
\end{eqnarray}
In Fig. \ref{f:fig3} we show the behavior of the overlap ${\cal O}$
and the conversion rate $\eta$ as a function of the scaled interaction time
for different intensity of the incoming twin-beam. A remarkable fact is
apparent: interaction times corresponding to high conversion rate also
optimize the overlap between the outgoing state and the theoretical PCS.
This means that the up-conversion, although only approximated, produces
a recombination process which is at the same time efficient and quite precise.
One should also mention that for the same interaction times one has
a small degree of mixing, indicating that the outgoing states are quite pure,
and minimum reciprocal peak likelihood, thus confirming good phase-coherence
properties. \par
\begin{figure}[h]
\psfig{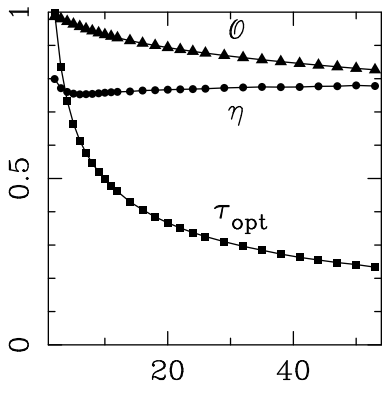}
\caption{The optimized overlap along with the corresponding
interaction time and conversion rate as a function of the twin-beam
input photons.}
\label{f:fig4}
\end{figure}

\noindent
In Fig. \ref{f:fig4} we report the maximum overlap, along with the
corresponding interaction time and conversion rate, as a function of the
twin-beam input energy $N_{in}=\langle\hat n_a + \hat n_b\rangle$.
The overlap ${\cal O}$ slowly decreases versus the input energy $N_{in}$,
whereas the conversion rate $\eta$ is almost independent on this quantity,
saturating to a value close to $80\%$. This results in a reliable
generation of  PCS with overlap between $80\%$ and $100\%$, for
outgoing states with energy $N_{out}=\langle\hat n_c\rangle$ up to
$N_{out}=20$ mean photon number.
The corresponding reciprocal peak likelihood $\delta\phi$ shows the scaling
$\delta\phi \propto N_{out}^{-3/4}$ which, though it is worse than the
ideal PCS performances $\delta\phi \propto N_{out}^{-1}$, it is far
superior to the coherent state level $\delta\phi \propto N_{out}^{-1/2}$. \par
The interaction time $\tau_{opt}$, which corresponds to the maximum overlap,
decreases with the input energy $N_{in}$. By a best fit on data in Fig.
\ref{f:fig4} we obtained the scaling power-law $\tau_{opt}
\simeq 1.4 N_{in}^{-0.45}$. Remarkably, the same scaling is observed as a
function of the output energy $N_{out}$, with only a
slight change in the proportionality constant
$\tau_{opt} \simeq 0.9 N_{out}^{-0.45}$.
\section{Conclusion}\label{s:outro}
In this paper we have suggested a scheme to generate the
phase-coherent states introduced in Ref. \cite{sha}.
The setup involves two $\chi^{(2)}$ nonlinear crystals and it is based
on parametric amplification of the vacuum followed by up-conversion of
the resulting twin-beam, the up-conversion playing the role of an
approximate photon number recombination. \par
We found that parametric approximation in down-conversion
of the vacuum state is valid also for moderate input power, and
that one has a considerably wide range of values of the interaction
time leading to an overlap very close to unit, the weaker is the
pump, the larger is this range. However, these values of
the interaction time correspond to a low conversion rate.
On the other hand, we found that the up-conversion process involved
in the second step of the scheme is both power-efficient and quite
precise in the generation of PCS. It is a remarkable fact
that the range of interaction times leading to high conversion rate also
optimize the overlap between the outgoing state and the theoretical PCS.
We have explored the case of twin-beam input photon number ranging from
$0$ to $54$, and we have observed a conversion rate about $80\%$,
with an overlap with ideal PCS between $80\%$ and $100\%$. This
corresponds to a reliable generation of PCS up to $N_{out}=20$ photons at
the output.
\section*{Acknowledgment}
M. G. A. Paris would like to acknowledge the ``Francesco Somaini''
foundation for partial support.
\vfill\eject

\vspace{80pt}
\par\noindent
\end{document}